# Quantifying Photoinduced Polaronic Distortions in Inorganic Lead Halide Perovskites Nanocrystals

Oliviero Cannelli[1], Nicola Colonna[2,3], Michele Puppin[1], Thomas Rossi[1,*], Dominik Kinschel[1], Ludmila Leroy[1,4], Janina Löffler[1], Anne Marie March[5], Gilles Doumy[5], Andre Al Haddad[5,†], Ming-Feng Tu[5], Yoshiaki Kumagai[5], Donald Walko[6], Grigory Smolentsev[7], Franziska Krieg[8,9], Simon C. Boehme[8,9], Maksym V. Kovalenko[8,9], Majed Chergui[1,§], Giulia F. Mancini[1,10§]

[1] École Polytechnique Fédérale de Lausanne, Laboratory of Ultrafast Spectroscopy (LSU) and Lausanne Centre for Ultrafast Science (LACUS), CH-1015 Lausanne, Switzerland.
[2] Laboratory for Neutron Scattering and Imaging, Paul Scherrer Institute, CH-5232 Villigen-PSI, Switzerland.
[3] National Centre for Computational Design and Discovery of Novel Materials (MARVEL), École Polytechnique Fédérale de Lausanne, CH-1015 Lausanne, Switzerland.
[4] LabCri, Universidade Federal de Minas Gerais, Belo Horizonte, Brazil.
[5] Chemical Sciences and Engineering Division, Argonne National Laboratory, 9700 S. Cass Ave., Lemont, IL 60439, USA.
[6] Advanced Photon Source, Argonne National Laboratory, 9700 S. Cass Ave, Lemont, IL, 60439, USA.
[7] Paul Scherrer Institute (PSI), 5232 Villigen, Switzerland.
[8] Institute of Inorganic Chemistry, Department of Chemistry and Applied Biosciences, ETH Zürich, Vladimir Prelog Weg 1, CH-8093 Zürich, Switzerland.
[9] Laboratory for Thin Films and Photovoltaics, Empa-Swiss Federal Laboratories for Materials Science and Technology, CH-8600 Dübendorf, Switzerland.
[10] Department of Physics, University of Pavia, I-27100 Pavia, Italy.
[§]Email: majed.chergui@epfl.ch; giuliafulvia.mancini@unipv.it

**Abstract**

The development of next generation perovskite-based optoelectronic devices relies critically on the understanding of the interaction between charge carriers and the polar lattice in out-of-equilibrium conditions. While it has become increasingly evident for $CsPbBr_3$ perovskites that the Pb-Br framework flexibility plays a key role in their light-activated functionality, the corresponding local structural rearrangement has not yet been unambiguously identified. In this work, we demonstrate that the photoinduced lattice changes in the system are due to a specific polaronic distortion, associated with the activation of a longitudinal optical phonon mode at 18 meV by electron-phonon coupling, and we quantify the associated structural changes with atomic-level precision. Key to this achievement is the combination of time-resolved and temperature-dependent studies at Br K-edge and Pb L$_3$-edge X-ray absorption with refined *ab-initio* simulations, which fully account for the screened core-hole final state effects on the X-ray absorption spectra. From the temporal kinetics, we show that carrier recombination reversibly unlocks the structural deformation at both Br and Pb sites. The comparison with the temperature-dependent XAS results rules out thermal effects as the primary source of distortion of the Pb-Br bonding motif during photoexcitation. Our work provides a comprehensive description of the CsPbBr$_3$ perovskites photophysics, offering novel insights on the light-induced response of the system and its exceptional optoelectronic properties.

---

[*] Now at: Department of Chemistry and Materials Research Laboratory, University of Illinois at Urbana-Champaign, Champaign, Illinois 61820, United States.
[†] Now at: Paul Scherrer Institute (PSI), 5232 Villigen, Switzerland.

**Introduction**

Lead halide perovskites are rapidly emerging as excellent candidates for optoelectronic applications, such as photovoltaics, light-emitting diodes [1], lasers [2], photodetectors [3] and quantum light sources [4], thanks to their outstanding performances and low fabrication costs [5]. These materials are characterized by facile processing routes, leading to defect-tolerant systems with widely tuneable band gaps, high photoluminescence (PL) quantum yields and narrow emission lines. Their potential stems from their extraordinarily long carrier lifetimes and diffusion lengths [6,7], which are in apparent contrast with previously reported low charge mobility[8] and lattice dynamical disorder [9].

The APbX$_3$ perovskite structure comprises a Pb-X (X=Cl$^-$,Br$^-$,I$^-$) inorganic framework made of flexible corner-sharing octahedra, with Pb$^{2+}$ cations surrounded by six halide anions, characterized by low-frequency phonons and a pronounced anharmonicity [10–14]. The A$^+$ cations, either inorganic (Cs$^+$) or organic (methylammonium, MA$^+$, or formamidinium, FA$^+$), fill the voids between PbX$_6$ octahedra. Distinct orthorhombic, tetragonal and cubic phases were identified in these systems, with phase transition temperatures varying with the cation composition [12,15–17]. In CsPbBr$_3$ nanocrystals (NCs), the phase diagram is characterized by a room temperature orthorhombic $Pnma$ crystalline group, with a transition to a tetragonal $P\frac{4}{m}bm$ group between 50 °C and 59 °C, and a higher temperature transition to a cubic $Pm\bar{3}m$ group between 108 °C and 117 °C[12].

Recent temperature-dependent studies showed competitive mechanisms underlying the thermal response in lead halide perovskites. Pair distribution function (PDF) analysis from X-ray powder diffraction in organic perovskites at room temperature showed significant internal local distortions of the PbX$_6$ octahedra [18]. The degree of these distortions was shown to increase with the temperature in MAPbBr$_3$ [19]. In CsPbX$_3$ NCs, structural defectiveness was revealed and ascribed to twin boundaries, whose density increases with temperature and leads to an apparent higher-symmetry structure that does, however, not correspond to the $Pm\bar{3}m$ cubic phase [11]. High energy resolution inelastic X-ray scattering and PDF studies on MAPbI$_3$ pointed to the presence of thermally-active anharmonic soft modes at 350 K [10], and local polar fluctuations among different non-cubic structures were confirmed in a low-frequency Raman study on MAPbBr$_3$ and CsPbBr$_3$ perovskites [14].

The peculiar lattice flexibility of lead halide perovskites also critically underpins their photodynamics. Time-resolved optical studies were conducted to understand key aspects of the early dynamics of the system [20–22]. Several works suggested the presence of polarons, *i.e.* charge carriers dressed by lattice distortions, in order to explain the time-resolved optical signals [13,23–26] and transport properties [27]. The polaron formation was shown to generally occur through the Fröhlich mechanism [28], which corresponds to a Coulomb interaction between the charge carriers and the macroscopic electric field created by the activation of longitudinal optical (LO) phonons [29,30].

Although the polaron hypothesis was frequently inferred to rationalize experimental observations in both organic and inorganic perovskites, the quantification of the associated local structural rearrangement is still missing. In hybrid organic-inorganic lead-halide perovskites, local distortions around the Pb [31] and Br [32] sites were separately reported in time-resolved X-ray absorption spectroscopy (TR-XAS) studies and ascribed to polaron formation, but an unambiguous identification of the associated structural fingerprint was not provided. Ultrafast electron diffraction on a MAPbI$_3$ thin film showed evidence of local rotational disorder of the PbI$_6$ octahedra arising from optical excitation, as a consequence of hot carrier-phonon coupling [33]. Only recently, was the presence of a dynamically expanding polaronic strain structurally identified in MAPbBr$_3$ single crystal with time-resolved diffusive X-ray scattering [34].

In all-inorganic lead halide perovskites the picture is still unclear. In a previous TR-XAS study of photoexcited CsPbBr$_3$ and CsPb(BrCl)$_3$ NCs at the Br K-, Pb L$_3$-, and Cs L$_2$-edges[35] carried out at a synchrotron, it was suggested that hole polarons form within the time resolution of the experiment (≤100 ps), around Br centres, with the halide ion turning into a neutral halogen, while electrons would be delocalized in the conduction band (CB). Additionally, the Cs sites did not show any response to photoexcitation. In a more recent time-resolved X-ray diffraction study (TR-XRD) at similar fluences, namely between 2.5 mJ/cm$^2$ and 12 mJ/cm$^2$, the lattice response upon photoexcitation was interpreted in terms of transient amorphization from a crystalline structure[36]. In both studies, however, the local structural distortion was invoked to rationalize experimental data in a qualitative way, rather than a quantitative one. A recent angle-resolved photoelectron spectroscopy (ARPES) study demonstrated an increase of the hole effective mass in CsPbBr$_3$ single crystals caused by electron-phonon coupling, and identified a specific LO phonon at 18 meV as the most coupled mode with the charge carriers [37].

In this work we demonstrate that, in CsPbBr$_3$ NCs, the 18 meV LO phonon mode is underpinning the structural distortion induced upon photoexcitation and we quantify the polaronic nuclear displacements with atomic precision. Specifically, we conducted a TR-XAS study at the Br K- *and* Pb L$_3$- absorption edges and we found that photoexcitation indeed induces polaron formation around Br centres, which also determines the response of Pb centres. We performed band structure calculations in which - to our knowledge for the first time in TR-XAS studies - the possible structural distortions are *a-priori* selected on a physical basis, fully accounting for core-hole final state effects on the XAS spectra. By comparing these accurate simulations to our experimental results, we identify the local PbX$_6$ octahedra bond distortions that contribute to the polaronic photoresponse. Additionally, we clarify the fundamental difference underlying optical and thermal activation in CsPbBr$_3$ NCs: our temperature-dependent XAS experiments result in quantitatively different spectral modifications compared to the optical activation, thereby excluding heat as the primary source of distortion of the Pb-Br bonding motif upon photoexcitation.

**Methods**

Br K-edge and Pb L$_3$-edge spectra have been recorded in the pre-edge and the XANES (X-ray absorption near edge structure) regions upon light or thermal activation of CsPbBr$_3$ NCs. The pre-edge region contains bound-bound core-to-valence transitions and is therefore sensitive to the density of unoccupied valence orbitals. The XANES includes the region just above the ionization limit (*i.e.* the edge). It is characterised by single and multiple scattering events of the photoelectron and, hence, it contains information about the bond distances and angles to the nearest-neighbour atoms around the probed site [38,39]. In its time-resolved implementation, the photoinduced changes of the TR-XAS spectrum reflect transient structural and electronic modifications at the probed sites and in their local environment [40].

The TR-XAS experiments were conducted at the 7ID-D beamline at the Advanced Photon Source (APS) of the Argonne National Laboratory [41,42]. A schematic representation of this experiment is shown in Figure 1. The sample consisted of long-chain zwitterion-capped $CsPbBr_3$ perovskite NCs with cuboidal shape (side length 10±2 nm) and high PL quantum yields [43]. Above band-gap photoexcitation was performed using a Duetto laser at a photon energy of 3.49 eV (*i.e.* 1.1 eV above the direct band gap excitation), a repetition rate of 1.304 MHz, 10 ps pulse duration and with a fluence of 8.8 $mJ/cm^2$, in the linear response regime. The photoinduced changes in the sample were probed at the Br K-edge (13.45-13.57 keV) and Pb $L_3$-edge (13.00-13.14 keV), with ~80 ps time resolution.

Comparative temperature-dependent Br K-edge and Pb $L_3$-edge static XAS was conducted at the SuperXAS beamline of the Swiss Light Source (SLS). The experiments were performed on dry $CsPbBr_3$ NCs enclosed in a thermostated cell holder. The thermal response of the system was monitored in the temperature range between 25 °C and 140 °C, where effects ascribed to either an increase in the NCs local structural disorder [11], or the occurrence of orthorhombic-tetragonal-cubic phase transitions [12,15] had previously been reported. Moreover, we acquired for each temperature step XRD patterns at 12.9 keV, below both absorption edges, to track longer-range structural changes and to assess the overall quality of the sample.

We performed accurate first-principles calculations using the Quantum ESPRESSO distribution [44,45], based on density functional theory (DFT) and plane-wave and pseudopotentials technique. The Perdew-Burke-Ernzerhof functional [46] was used to describe electronic exchange-correlation effects. The electron-ion interaction was modelled using ultrasoft pseudopotentials from the PS-library [47]. The projected density of states (p-DOS) was computed across the band gap. XANES Br K-edge spectra were simulated with the XSpectra code [48,49] of Quantum ESPRESSO, explicitly accounting for the screened core-hole effect in separate supercell calculations for each non-equivalent Br atom, and calculating the average

Br K-edge spectra. XANES Pb L$_3$-edge calculations were not carried out due to the limitations in the explicit inclusion of a screened core-hole in describing holes with non-zero orbital momentum [50], as in the case of the Pb 2p$_{3/2}$ orbital. Details about all experimental methods, the data acquisition scheme, and the computational methods and DOS calculations, are described in the Supplementary Information (SI).

**Results**

*TR-XAS*

The steady-state Br K-edge and Pb L$_3$-edge spectra, normalized to the last data point of the post-edge region, are shown in Figures 2a and 2b (black solid line). Our calculations of the p-DOS show that the top of the valence band (VB) is composed of Br 4p orbitals, with a non-negligible proportion of Pb 6s orbitals and a minor contribution of Pb 6p orbitals, while the CB is largely dominated by the Pb 6p orbitals (SI, Fig. S3).

The photoinduced changes are reflected in the transient spectra, defined as the difference of the excited minus un-excited XAS spectra, and shown for UV pump/X-ray probe time delays of 100 ps (red), 10.1 ns (yellow) and 163.5 ns (grey). The Br transient spectra (Fig. 2a) were scaled by the inverse of the absolute area underlying the curves, i.e. x155 (100 ps), x363 (10.1 ns), x976 (163.5 ns). The same scaling factors were used for the Pb transients (Fig. 2b, details in the SI). Notably, even though the amplitude of the TR-XAS decays over time, the profiles of both Br and Pb transient spectra remain unchanged.

The Br K-edge transients show prominent peaks at the pre-edge (13.4675 keV), main-edge (13.472 keV) and post-edge (13.4765 keV) regions. The first feature was ascribed to the opening of a new 1s-4p channel following the creation of holes in the VB upon photoexcitation [35]. Such a scenario also implies a blue shift of the edge and, indeed, the second and third features could partially be reproduced in the difference spectrum of the blue-shifted ground state spectrum minus the unshifted one. However, this qualitative approach does not account for all the modulations that show up in the above-edge region, which generally point to photoinduced structural changes. We shall address these later, using *ab-initio* calculations, and demonstrating their connection with photoinduced structural changes. The Pb L$_3$-edge steady-state spectrum exhibits featureless edge and XANES regions. The transients are characterized by two positive features in the pre-edge region (13.031 keV and 13.038 keV), a negative peak at the edge position (13.043 keV) and a positive peak in the post-edge region (13.060 keV). The reduction of Pb centres upon photoexcitation of the electrons in the CB, which is mainly composed by Pb p-orbitals, is not compatible with the transient line shape, as discussed by Santomauro *et al.* [35]. The appearance of pre-edge features in the transients traces can only be explained by the opening of new channels from the 2p core orbitals. Indeed, due to hybridization, depleting the VB not only affects the Br centres but also the Pb ones, according to the computed p-DOS (Figure S3). Core-to-valence transitions can occur into the Pb 6s orbitals, which have a non-negligible contribution towards the top of the VB. Above the edge, the transient features are due to photoinduced structural changes. Because the Pb atoms are affected by the structural distortion around the Br centres (see below), it is likely that the above-edge features of the Pb $L_3$ absorption transient in part reflect the latter.

The decay kinetics at both Br and Pb main edges are shown in Figure 2c. The traces were normalized to their maximum value, allowing a straightforward comparison of the time-resolved signal of both centers. It is clear that both traces show the same temporal evolution within the noise level. The data were analyzed following a global fit procedure for both traces. The best fit results were obtained with a bi-exponential decay function and a flat offset, which persists up to the time limit explored in our time traces (130 ns). The fit function was convoluted with a Gaussian profile ($\sigma$=45 ps), representing the instrument response function of our measurements. The recorded time constants (pre-exponential factors) are $\tau_1$=120±20 ps (60%), $\tau_2$=900±300 ps (21%) and an offset (infinite times of 19% amplitude). Specifically, the fast time component $\tau_1$ is compatible with Auger recombination, where an electron in the CB and a hole in the VB recombine, in a non-radiative process, transferring their energy to a third carrier. Supporting this interpretation, recent fluence-dependent PL and transient absorption studies on $CsPbBr_3$ NCs reported Auger recombination acting on this time scale[51,52]. $\tau_2$ is ascribed to the radiative recombination of the photoexcited charge carriers, *i.e.* holes from the VB and electrons from the CB, in general agreement with PL studies [43,53].

*Thermal XAS*

Given the on-going debate about photoinduced electronic and thermal effects [36] and considering that our pump pulse deposits an excess energy of the photocarriers of ~1 eV, it is important to disentangle electronic from thermal effects in the photoinduced response presented here.

In our pump-probe experiment, the hot carriers generated by the pumping process dissipate their excess energy through carrier thermalization in the sub-100 fs regime [20] and, immediately after, by charge carrier cooling on sub-ps time scales [54]. These events determine impulsive heating of the crystalline lattice. If the energy deposited on the system is sufficiently high, therefore, this process might translate into impulsively activated orthorhombic-tetragonal-cubic phase transitions [12]. At later time scales, the hot lattice relaxes through heat transfer to the solvent and/or the ligands. In ligand-stabilized colloidal NCs in solution this process should be completed in sub-ns time scales [55], due to the efficient vibrational coupling between the NC, the ligands and the solvent molecules. Our TR-XAS experiment looks at the system relaxation in time scales from 80 ps onwards after photo-excitation, *i.e.* when the thermal equilibration of the lattice with the surrounding bath has already initiated. At these time scales, the photoinduced relaxation of the system and its purely thermal and temperature-dependent responses can be directly compared. This assumption can be harnessed to verify whether the optically-induced relaxation coincides with thermodynamic lattice cooling.

Figures 3(a, b) show the thermal difference spectra at the Br K and the Pb $L_3$ edges (full thermal spectra are available in the SI). These are obtained by subtracting the 25 °C spectrum from the T-dependent XANES spectra. Figures 3(c, d) show the unscaled pump-probe difference spectra at 100 ps (red), 10.1 ns (yellow) and 163.5 ns (grey) for each absorption edge. The thermal difference spectra at the Br K-edge (Fig. 3a) display an intensity change through the thermal gradient, with an overall area decrease in the energy range 13.466–13.478 keV lowering the temperature from 140 °C to 25 °C. Particularly, the features at 13.468 keV and 13.472 keV have two different temperature dependences, the former becoming dominant for temperatures below 65 °C. Main differences between photoinduced and thermal data sets can be found in the Br K pre-edge region. Specifically, the negative feature at 13.468 keV found in the thermal data set is absent in the pump-probe spectra, which instead are characterized by a positive peak centered around 13.4675 keV. In the case of Pb, thermal difference spectra in Figure 3b display a broad negative feature covering the 13.035-13.053 keV spectral range, whose intensity uniformly decreases lowering the temperature, and a single pre-edge feature centred at 13.032 keV, with intensity increasing upon temperature decrease. In contrast and as already discussed, the pump-probe Pb spectra

show two pre-edge peaks, respectively at 13.031 keV e 13.038 keV, whose intensity decreases as a function of time delay.

To quantify overall spectral changes as a function of either temperature or optical excitation, at both edges, we introduce a comparison parameter *C* defined as:

$$C = \sum_{E_i} I(E_i) / \sum_{E_j} I(E_j)$$

with $I(E_i)$ and $I(E_j)$ being the spectral intensities at the energy points $E_i$, $E_j$ in the main-edge and pre-edge regions, respectively. In Figures 3 a-d (grey areas), $E_i$(Br) = [13.471-13.472] keV; $E_j$(Br) = [13.468-13.469] keV; $E_i$(Pb) = [13.044-13.047] keV; $E_j$(Pb) = [13.037-13.040] keV. The indicated intervals comprise equally-spaced energy points. Figures 3(e, f) show the parameter *C* for Br and Pb, respectively: it expresses the relative intensity ratio between two spectral regions of the same data set, and it describes the entity of XAS spectral shape changes through a temperature gradient (purple dots) or upon photoexcitation (black crosses).

Within the error bars, *C* values are constant in the pump-probe case for either Br and Pb, in agreement with the spectral evolution of the TR-XAS signal. Indeed, we observe a synchronous systems's response throughout the TR-XAS spectrum in its decay to the ground state. We remark that negative values of *C* are due to the presence of the pre-edge and main-edge features that have opposite signs, respectively at the energies 13.4675 keV and 13.472 keV for the Br K-edge, and at 13.038 keV and 13.044 keV for the Pb $L_3$-edge. Starting from 140 °C, the *C* parameter for the Br thermal differences has stable positive values for temperatures down to 60 °C, and undergoes a progressive change from positive to negative values lowering the temperature in the interval 60-35 °C, due to line shape modifications in the edge region. Instead, Pb thermal differences show positive *C* values at all temperatures, which originate from the negative sign of pre-edge and main-edge features. Since both features decrease in amplitude upon temperature decrease, Pb *C* values remain essentially unchanged with the thermal gradient.

Based on the radically different behaviour of the *C* parameter for the optical and thermal data sets, we can safely conclude that the photoresponses at the Pb and Br edges reported here are not affected by thermal effects, which likely occur on shorter time scales than our temporal window, as discussed in the following. Thus, we rule out the hypothesis that the photoexcited state corresponds to thermally-driven lattice changes [36].

**Theoretical simulations: photoinduced polaronic distortion**

The XAS spectrum reflects the probability of an electronic excitation from a core orbital to the unoccupied states of the system which lies at higher energies than the Fermi level. Hence, the first-principles description of XANES spectra requires the computation of highly-localized initial orbitals and of the unoccupied conduction states, the latter in the presence of a screened core-hole, since it reflects the possible final states with a missing core electron.

In condensed matter systems, this level of accuracy is retrieved relying on band structure calculations, where the effect of the screened core-hole is explicitly accounted for as, *e.g.*, in supercell simulations [56,57]. Band structure calculations also allow access to the phonon spectrum of the system, the atomic displacements occurring upon phonon mode activation, and the electron-phonon coupling between the charge carriers and the lattice degrees of freedom. To date XAS spectra of several solid state systems have been computed using *ab-initio* methods, which rely on real-space atomic clusters, especially for the simulation of excited state spectra [31,58–60]. In these cases, the interest was focused on determining the local structural distortions in the photoexcited system, as in the case of charge carriers trapping, rather than understanding the origin of its structural response in the presence of charge carriers, which requires electron-phonon coupling calculations.

First principles computations of pump-probe spectra are generally simulated either using *a-posteriori* strategies, *i.e.* gradually modifying the local structure of a small atomic cluster until the best agreement between the simulation and the experiment is achieved [31,58,60], or selecting *a-priori* specific subsets of configurations, where stricter constraints are imposed on a physical basis [59]. Here, to our knowledge for the first time in TR-XAS, we adopt an approach based on *ab-initio* calculations performed under periodic boundary conditions with an *a-priori* selection of the structurally distorted states. Consistent with the phase diagram of $CsPbBr_3$ perovskite [12,15], we computed the ground state considering the atomic positions of the *Pnma* orthorhombic cell, as derived from room-temperature XRD[15]. The lattice perturbation caused by the optical pump was then simulated following two possible scenarios.

First we consider the scenario, ruled out experimentally, in which a thermally-induced phase transition to an ordered $Pm\bar{3}m$ cubic state might occur. In fact, we estimated an upper limit of ΔT~120 °C to the impulsive heating generated by our pump pulse (see SI) which, in the absence of lattice cooling and consistent with the CsPbBr$_3$ phase diagram, would lead the system from the orthorhombic to the cubic phase. For our simulation we use the atomic coordinates available in the literature [15]. Second, we consider a polaronic distortion induced by electron-hole pair excitation, introducing a structural modification along the 18 meV phonon mode which is the most strongly coupled to the charge carriers *via* electron-phonon coupling [37]. The structural modification caused by the aforementioned process is schematized in Figure 4a. Key to this method is the adoption of a band structure calculation, which allows to: (i) identify the phonon mode with the strongest electron-phonon coupling and; (ii) to introduce the phonon distortion in the periodic lattice, thereby approximating the large polaron spatial extension over multiple unit cells, in agreement with the literature [13,37]. Electronic non-equilibrium effects following the optical excitation were not included in the calculations, being too computationally intensive in the presence of an explicit core-hole description.

Figure 4b compares the simulated Br K-edge absorption spectra for the orthorhombic ground state (red solid) and the cubic state (blue dots) with the experimental steady-state spectrum (black solid). The three X-ray absorption spectra are characterized by a first peak arising from the Br 1s-4p electronic transition around 13.472 keV. The above-edge spectral modulations for the calculated orthorhombic ground state best reproduce the experiment, as expected from the CsPbBr$_3$ perovskite phase diagram, which is characterized by a *Pnma* orthorhombic symmetry at room temperature. The simulated spectrum for the cubic phase shows a modulation mismatch with respect to the experiment which is prominent in the 13.492-13.505 keV energy range.

Figure 4c zooms into the above-edge region beyond 13.480 keV of the experimental pumped and unpumped spectra (respectively, yellow and black traces), and for the simulated orthorhombic ground state and the polaronic distorted state (respectively, red and dashed-green traces). Even though the photoinduced changes are more pronounced in the experiment, the simulation faithfully follows the photo-induced spectral modification, with intensity depletions of the 13.485 keV and 13.508 keV maxima and an intensity increase of the 13.497 keV minimum.

Figure 4d shows the simulated pump-probe signals obtained subtracting the XAS spectrum of the orthorhombic ground state from the XAS spectrum due to the polaronic distortion (red solid) as well as from the cubic phase XAS spectrum (blue dots). The comparison of the difference curves allows to remove possible systematic errors of our *ab-initio* calculation for both ground and excited states. Above the edge, the experimental pump-probe spectrum at 100 ps shows very good agreement in both position and relative amplitude assuming a polaronic lattice distortion generated by the optical activation of the 18 meV LO phonon within the polar inorganic lattice. On the other hand, there is a clear disagreement with the simulation that assumes the phase transition to the $Pm\bar{3}m$ cubic structure. We highlight that the XAS simulation for the orthorhombic-cubic phase transition does not reproduce the spectral line shape of the 120 °C minus 25 °C thermal difference reported in Figure 3a either. This result is analyzed in a separate work.

We remark that the residual deviations between the simulated polaronic distortion and the experimental pump-probe spectra (Fig. 4d) can be rationalized considering that electronic effects caused by the optical pump are absent in the calculation. Indeed, the edge region is particularly sensitive to photoinduced changes of the unoccupied DOS of the system. Relying on the one-electron approximation, in Br K-edge transitions the initial and final states differ by the presence of a core-hole in a 1s Br orbital and a photoelectron above the Fermi level. Due to the localization of the 1s orbitals on Br atoms, the Br K-edge transition probability is non-negligible only for final states where the photoelectron has a significant character of the Br absorbing atom, which are present both in the VB and CB (see Fig. S3). When the band occupancy is perturbed by the optical pump, the Br XANES in the edge-region is also modified.

**Discussion**

The comparison of optical XAS study with temperature-dependent XAS measurements rules out a dominant photoinduced thermal effect in the TR-XAS response. Indeed, if the evolution of the pump-probe spectral line shape reflected the lattice cooling following impulsive heating, a change in the TR-XAS signal intensity and line shape similar to Figures 3(a, b) should be expected. However, this is not observed in the transients reported in Figures 3(c, d). Furthermore, this discrepancy is confirmed by the differences in the *C* parameters of Figures 3(e, f) between the thermal and optical data sets. Thus, even though a significant heat deposition could occur under our experimental conditions (see SI for calculations), a thermal origin of the transient signal can hardly justify the strong difference between the time evolution of the pump-probe signal and the changes expected for thermal cooling.

In ligand-stabilized colloidal NCs, heat transport is known to be determined by the organic/inorganic interface rather than the thermal conductivity of the inorganic core of the NC [61]. In CdSe NC systems, the heat loss from the ligand-NC complex to the bath was observed in 150-320 ps, depending on the solvent [55]. Considering the similarity of the ligand-capped CdSe system and our NCs in size, ligand composition, low-energy NC phonon spectrum, fast thermalization dynamics [55,62], and the main role of the ligand-solvent coupling to the cooling process, analogous time scales are expected for our $CsPbBr_3$ NCs in solution. Relying on Newton's law to describe the NC lattice cooling and assuming a relatively slow $\tau_{cooling}$ of 300 ps, an initial $\Delta T \sim 120$ °C temperature would quickly drop to smaller values, e.g. $T_{NC}(t=600\text{ ps}) \sim 40$ °C.

This prediction is in stark contrast with the persistency of the TR-XAS signal over time, which preserves the same line shape at the Br and Pb edges up to the longest time delay measured in our pump-probe experiment, namely 163.5 ns. We conclude that heat dissipation in zwitterion-capped $CsPbBr_3$ perovskite NCs should be complete in shorter time scales than our TR-XAS time resolution, not affecting the pump-probe measurements. Notably, similar results were reported in a Pb $L_3$-edge TR-XAS investigation on $MAPbBr_3$ ligand-capped NCs in solution [31], where the significant heat load caused by the pump energy deposition into the NC lattice was argued to be dissipated in time scales shorter than 100 ps.

The light-activated structural modification is not compatible with a cubic crystalline structure, nor is it due to disorder, amorphization, or melting caused by thermal effects. Our theoretical analysis clarifies key aspects of the photoinduced response of $CsPbBr_3$ perovskite NCs, ascribing the excited state structural changes to the presence of distinct polaronic distortions that the XAS simulation specifically identifies. Indeed, the atomic displacements of the Pb-Br framework are traced to the distortion of the 18 meV LO phonon mode, which is the most strongly responsive to the charge carriers *via* electron-phonon coupling. Strong electron-phonon coupling in lead-halide perovskites was demonstrated in the electronic structure of $CsPbBr_3$ single crystals, where signatures of large hole polarons were identified by ARPES and attributed to the activation of the same LO phonon mode [37]. Moreover, previous time-domain results based on the optical Kerr effect [13], electronic resonant and non-resonant impulsive vibrational spectroscopy [24], ultrafast THz studies [26] and 2D electronic spectroscopy [23], were rationalized in terms of polaron formation in organic and inorganic perovskites. Polaronic strain was also invoked as the primary driving force of light-induced phase separation in multi-halide perovskites, explaining the reversibility of the process, its dependence on the number of photocarriers and the self-limiting size of the domains[63]. The importance of electron-phonon coupling on the $CsPbBr_3$ electronic response was also confirmed by PL investigations [64] and time-resolved 2D electronic spectroscopy [25], pointing to a relevant influence of LO phonon modes with energies between 16 meV and 19 meV, and attributed to the lead-halide inorganic framework, consistently with our findings.

Thanks to the agreement with the TR-XAS exprimental results, our simulations provide a compelling atomic-level description of the polaronic distortion. As depicted in Figure 4a, the distortion along the 18 meV phonon mode implies that the Pb-Br bonds are asymmetrically shortened along the orthorhombic *c*-axis, moving Pb cations out of the octahedral inversion centre, and substantially displacing the axial Br nuclei from their equilibrium position, whereas the equatorial Br centres and the Cs ions are marginally affected. Specifically, the photoinduced displacement of the axial Br atoms along the *c*-axis (Pb-$Br_{axial}$ equilibrium bond distance = 2.958 Å) is 6 times larger than the equatorial Br atoms (Pb-$Br_{equatorial}$ equilibrium bond distance = 2.964 Å) and 2.5 times more pronounced than the Pb off-centre displacement. The post-edge modulations observed in the Pb $L_3$-edge transients can thus be explained by the displacements of the Pb and Br centres caused by the photogenerated polaronic distortion. Notably, the absence of a photoinduced structural response from the Cs centres reported in [35] also agrees with this finding. Indeed, in all-inorganic perovskites the $A^+$ cation allocated in the lattice cuboctahedral voids is largely mobile and its dynamics is essentially decoupled from the inorganic Pb-X framework[9]. Further supporting the above description is that the kinetic traces and transient XAS energy profiles point to a concerted behaviour of Br and Pb in response to the optical excitation. The time scales of the intensity decays are fully in line with the Auger [51,52] and PL recombination lifetimes [43,53] in $CsPbBr_3$ perovskites. The relaxation occurs with a direct recovery of the perovskite's ground state, as confirmed by the retention of the TR-XAS line shapes in the decay process and by the time evolution of the *C* parameter.

The high PL quantum yields reported for CsPbBr$_3$ NCs[43] and the fact that the transients do not change line shape profile, *i.e.* there is no evidence for intermediate states, point to a recovery of the system largely dominated by charge carrier recombination. The presence of polaronic distortions is consistent with this scenario: after the photocarriers have induced the lattice displacements dictated by the strong electron-phonon coupling of the system, the subsequent electron-hole recombination causes the reversible unlocking of the structural distortions of the Pb-Br framework, back to the ground state configuration.

Notably, spectral line changes as a function of time delay were detected in Cs$_3$Bi$_2$Br$_9$ perovskites with Br K-edge TR-XAS as a consequence of their asynchronous electronic and structural relaxation upon optical excitation [58], with long-persisting lattice disorder after charge carrier recombination. The observation of short-lived valence holes in Cs$_3$Bi$_2$Br$_9$, compared to the post-edge signatures of lattice distortion, indicates that composition and structure of the inorganic sublattice in halide perovskites, either Bi$_2$Br$_9$ or PbBr$_3$, can strongly influence the photodynamics of the system and thus its optoelectronic performances.

This work also highlights the importance of local structural techniques in unraveling the nature of electronic and structural changes in perovskites, triggered by different external perturbations. In diffraction, structural modifications are obtained using approaches that go beyond standard Rietveld refinement methods. One of these methods relies, for example, on the computation of the Fourier transform of the total scattering structure factor to retrieve the PDF, which expresses a probability of finding pairs of atoms separated by a distance *r* [65]. PDF analysis from X-ray powder diffraction in a host of organic perovskites showed significant internal local distortions of the octahedra at room temperature [18]. Later, total scattering structural characterization, relying on a joint Debye scattering equation/atomic PDF approach, clarified that in CsPbBr$_3$ NCs the structural defectiveness is due to twin boundaries, whose density increase with temperature leads to an apparent higher-symmetry structure that does however not correspond to the $Pm\bar{3}m$ cubic phase [11]. A recent high energy resolution inelastic X-ray scattering and PDF study on MAPbI$_3$ pointed to the presence of thermally-active anharmonic soft modes at 350 K, corresponding to in-phase and out-of-phase rotations of the PbI$_6$ octahedra [10]. Shortly after, local polar fluctuations were also confirmed in MAPbBr$_3$ and CsPbBr$_3$ perovskites in a temperature-dependent Raman study, where the presence of a zero-frequency Raman peak was assigned to anharmonic thermal fluctuations among different non-cubic structures [14].

All these studies underline that correlating medium-to-long range structural methods with local probes helps distinguishing subtle changes in the perovskite lattice. In this respect, XAS represents a correlative short-range structural tool to probe disordered or dynamically changing systems such as lead halide perovskite NCs. In its time-resolved implementation, TR-XAS offers the advantage of combining electronic and local structural sensitivity, making it an ideal technique to probe lattice modifications induced by the presence of photocarriers, as in the case of polaron formation [31,32,35,58] or charge trapping [59,66], and to discern them from thermally-induced changes.

**Conclusions**

We presented results of light- and temperature-induced changes at the Br K-edge and Pb L$_3$-edge of CsPbBr$_3$ NCs dispersed in toluene solution or as dry powders. Our results show strong differences between the thermal and optical response of the system, excluding dominant photothermal effects in the observed pump-probe dynamics. The photoinduced spectral changes at the Br K-edge, stemming from a polaron distortion, are here quantified for the first time using advanced band structure calculation, including an *a-priori* selection of the excited state and fully accounting for core-hole effects on the TR-XAS spectra. The comparison between our experiment and theory identifies the lattice changes at the origin of the transient Br post-edge modulations with a distortion along a LO phonon mode at 18 meV. These simulations represent an atomic-level description of the light-induced nuclear displacement, dominated by an asymmetric Pb-Br bond shortening along the orthorhombic *c*-axis. This is supported by the identical kinetic evolution of the transient Br K-edge and Pb L$_3$-edge transients, which show that the latter is a direct consequence of the polaronic distortion around Br centres. This is also consistent with the high PL quantum yields reported for CsPbBr$_3$ NCs [43] and provides new microscopic insights in the Pb-Br sublattice dynamics, clarifying the perovskite response under light-induced out-of-equilibrium conditions.

**Data Availability**

Processed data showed in this manuscript are available in the Supplementary Information. Raw Br K-edge and Pb $L_3$-edge XAS data were generated at APS and SLS large-scale facilities and are available in the repository: http://doi.org/10.5281/zenodo.4564629

**Acknowledgements**

This work was supported by the European Research Council Advanced Grant H2020 ERCEA 695197 DYNAMOX and by the SwissNSF NCCR-MUST and NCCR-MARVEL. GFM acknowledges the support of the European Union's Horizon 2020 research and innovation programme (grant agreement No. 851154). The Argonne group (GD, AMM, AA, YK, M-FT) was supported by the U.S. Department of Energy, Office of Science, Basic Energy Sciences, Chemical Sciences, Geosciences, and Biosciences Division under contract DE-AC02-06CH11357. This research used resources of the Advanced Photon Source, a U.S. Department of Energy (DOE) Office of Science User Facility operated for the DOE Office of Science by Argonne National Laboratory under Contract No. DE-AC02-06CH11357. We thank Balázs Őrley for the graphical rendering of Figure 1.

**Author contributions**

MC conceived the experiment and designed the experiment. OC, TR, DK, AMM, GD, AA, M.-FT, YK, DW performed the APS experiment, OC, LL, GS, GFM performed the SLS experiments. FK and MVK provided the samples. NC performed all DFT and XCH simulations. OC, AMM, GFM analysed the data. OC, NC, MP, MC, GFM interpreted the data. OC, MC and GFM wrote the manuscript. All authors read and contributed to the manuscript.

**Additional Information**

Supplementary information: (1) Samples and characterization. (2) Experimental methods. (3) Fluence scans data analysis. (4) Energy scans data analysis. (5) Time scans data analysis. (6) Estimation of the number of photocarriers. (7) Computational methods and DOS calculation. (8) T-dependent XRD and XAS.

**Competing Financial Interest**

The authors declare no competing financial interests.

**Figures**

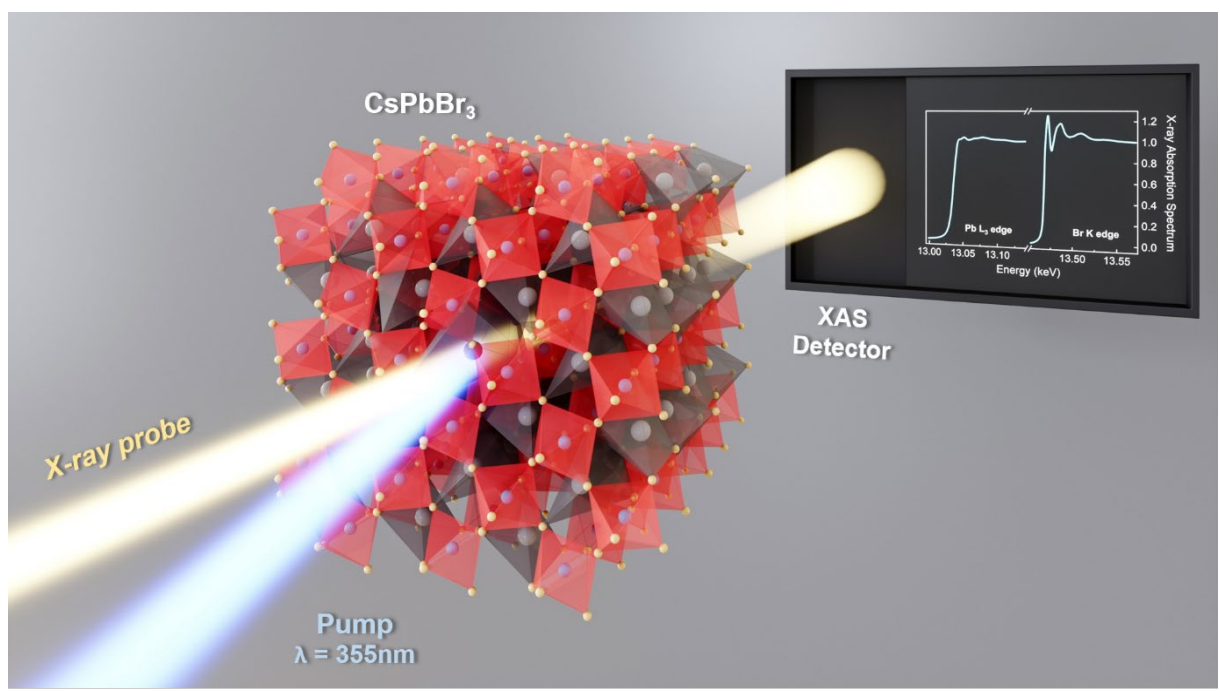

**Figure 1 | Ultrafast element-selective probing of optically-induced polaronic distortions in CsPbBr$_3$ perovskite NCs.** Schematic layout of the experiment. TR-XAS was conducted on long-chain zwitterion-capped CsPbBr$_3$ NCs dispersed in toluene solution with a concentration of 5.8 mg/ml and flowed through a flat jet. The laser pump (355 nm) and the X-ray probe, Br K-edge (13.450-13.570 keV) and Pb L$_3$-edge (13.000-13.140 keV), were in almost collinear geometry.

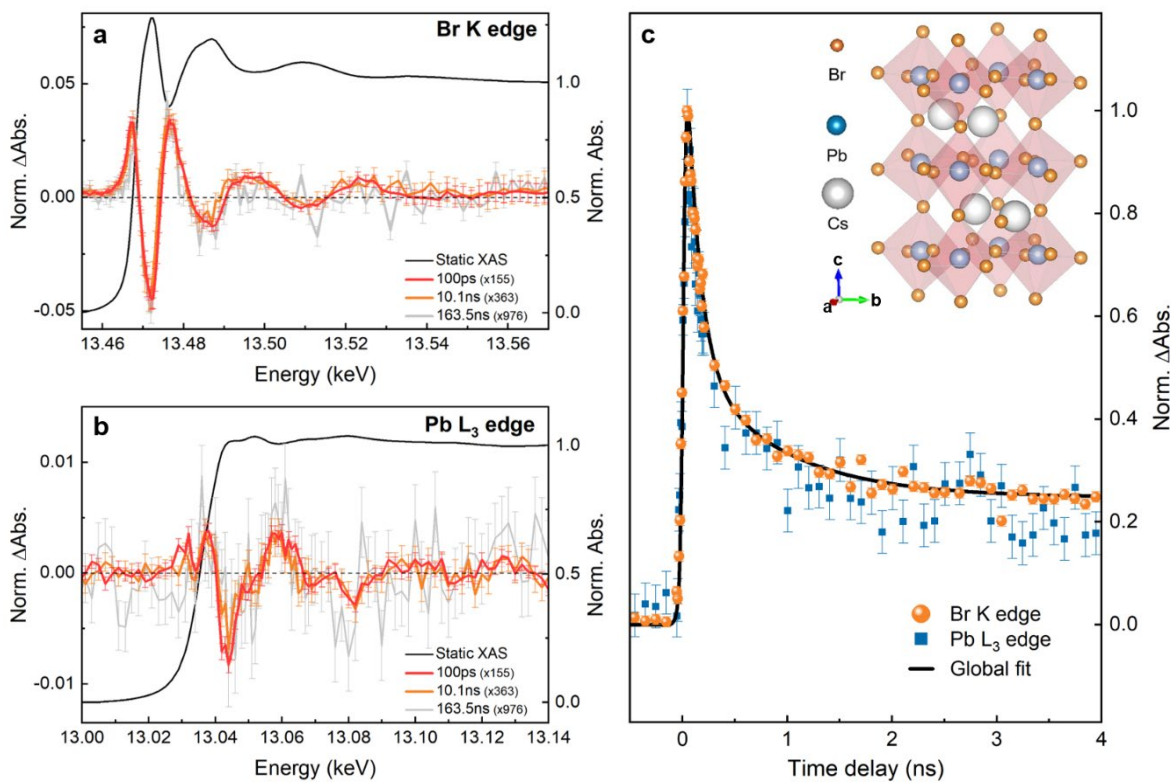

**Figure 2 | X-ray Absorption Energy and Time Traces.** (**a**) Br K-edge and (**b**) Pb L$_3$-edge XAS spectra: steady-state (black) and energy transients at 100 ps (red, x155), 10.1 ns (yellow, x363), 163.5 ns (grey, x976) time delays. The error bars correspond to the standard error of the measurements. (**c**) TR-XAS time traces at the Br K-edge (13.472 keV, orange), Pb L$_3$-edge (13.043 keV, light-blue) and the exponential fit (black). The error bars were computed as the error propagation of the pumped and unpumped scans, calculated as the square root of the total single photon counts. Inset: a graphical representation of $Pnma$ orthorhombic CsPbBr$_3$ [67]. The Br, Pb and Cs atoms are respectively reported as orange, light-blue and grey spheres.

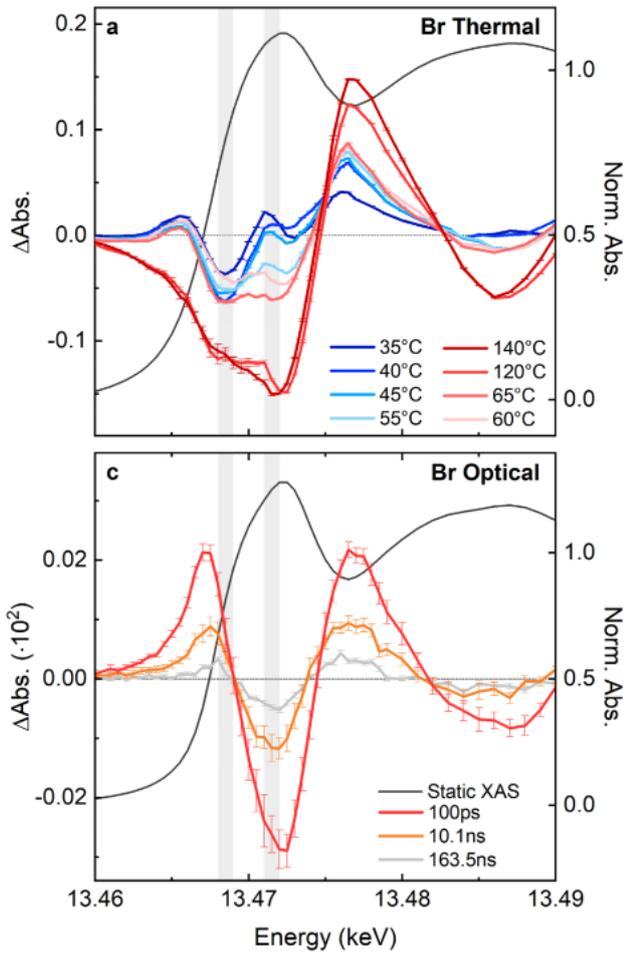
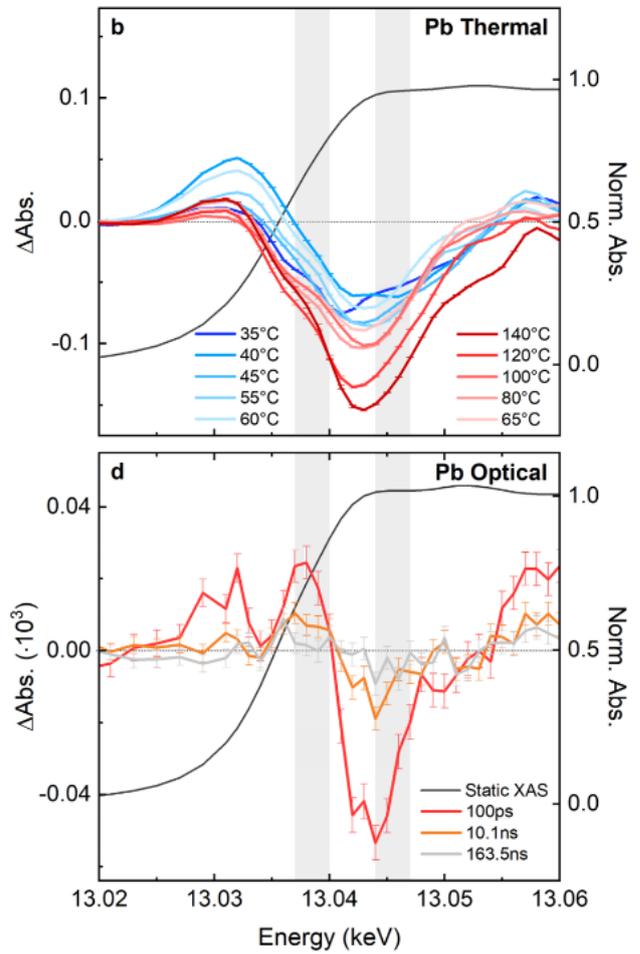
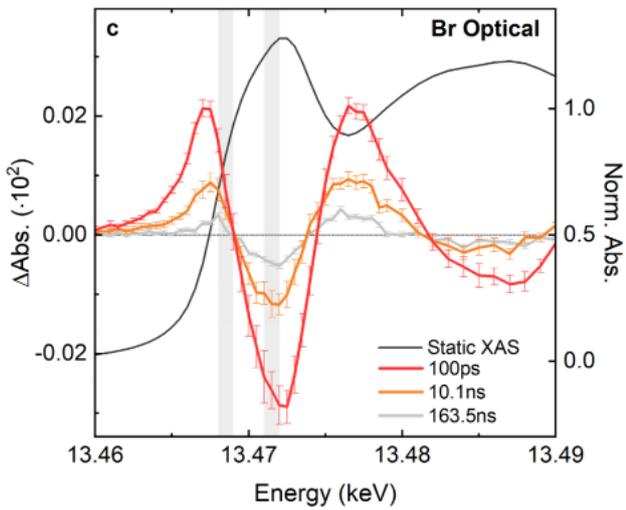
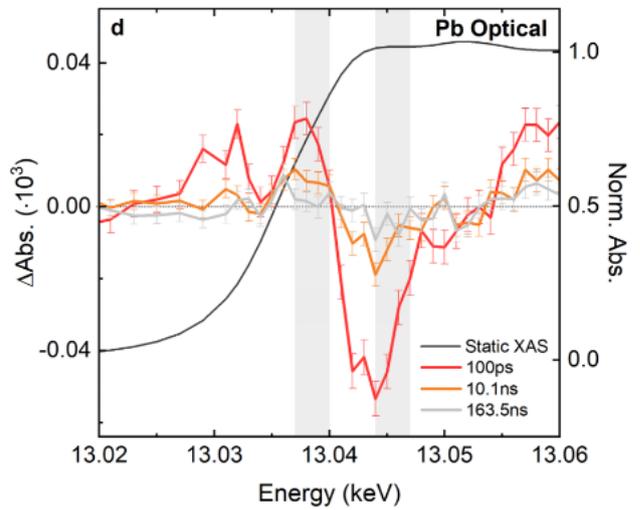
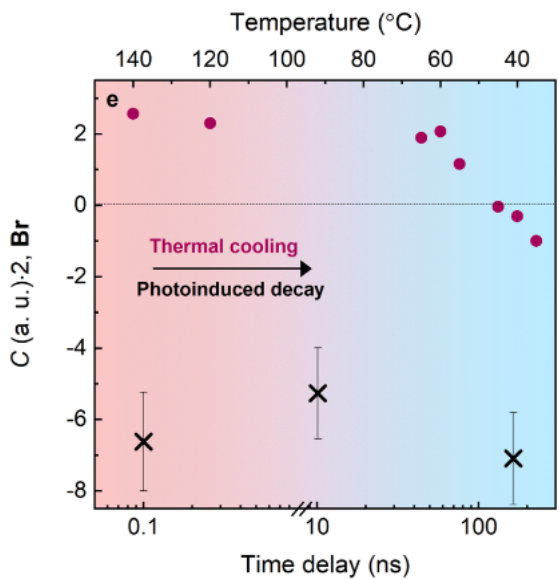
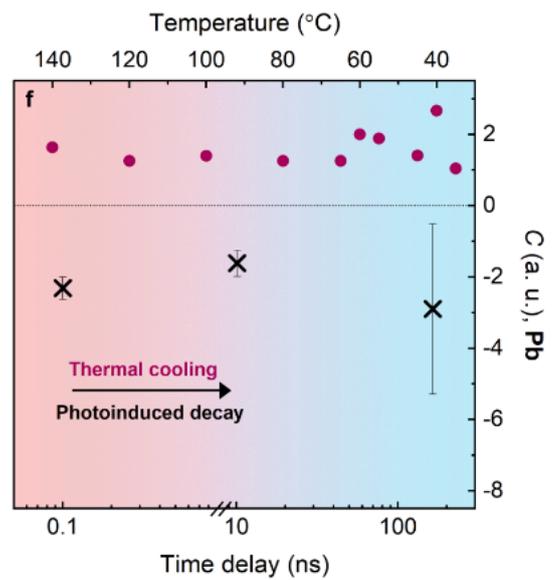

**Figure 3 | Comparison between photoinduced and thermally-activated XAS transition in CsPbBr$_3$ at the Br K-edge and Pb L$_3$-edge**. (**a**) Br K-edge steady-state (dark grey) and temperature-dependent XAS differences from 35 °C to 140 °C (respectively from blue to red) and (**b**) Pb L$_3$-edge steady-state (dark grey) and temperature-dependent XAS differences from 35 °C to 140 °C (respectively from blue to red). In both panels, the XAS differences were computed by subtracting the 25 °C spectrum from the temperature-dependent XAS spectra. Before performing the differences, all steady-state temperature-dependent spectra were baseline corrected and scaled by their underlying areas (consistently with the data treatment of time-resolved spectra). A 3-point adjacent averaging of the spectra was performed to better track the evolution of the spectral shape as a function of the temperature. (**c**) Br K-edge steady-state (dark grey) and pump-probe spectra at 100 ps (red), 10.1 ns (yellow) and 163.5 ns (grey) and (**d**) Pb L$_3$-edge steady-state (dark grey) and pump-probe spectra at 100 ps (red), 10.1 ns (yellow) and 163.5 ns (grey). The steady-state spectra in panels (a),(c) and (b),(d) show the same spectral shapes, accounting for their different energy resolution. (**e**) Br K-edge comparison parameter *C* as a function of the temperature (purple dots, top axis) and pump-probe time delay (black crosses, bottom axis), defined as the ratio of the averaged XAS difference in the energy interval 13.471-13.472 keV and 13.468-13.469 keV, corresponding to the shaded areas in panels (a),(c). (**f**) Pb L$_3$-edge comparison parameter *C* as a function of the temperature (purple dots, top axis) and pump-probe time delay (black crosses, bottom axis), defined as the ratio of the averaged XAS difference in the energy interval 13.044-13.047 keV and 13.037-13.040 keV, corresponding to the shaded areas in panels (b),(d). Br *C* values for both thermal and optical data sets were multiplied by a factor x2 in order to enable a straightforward comparison between Br and Pb results in panels (e),(f).

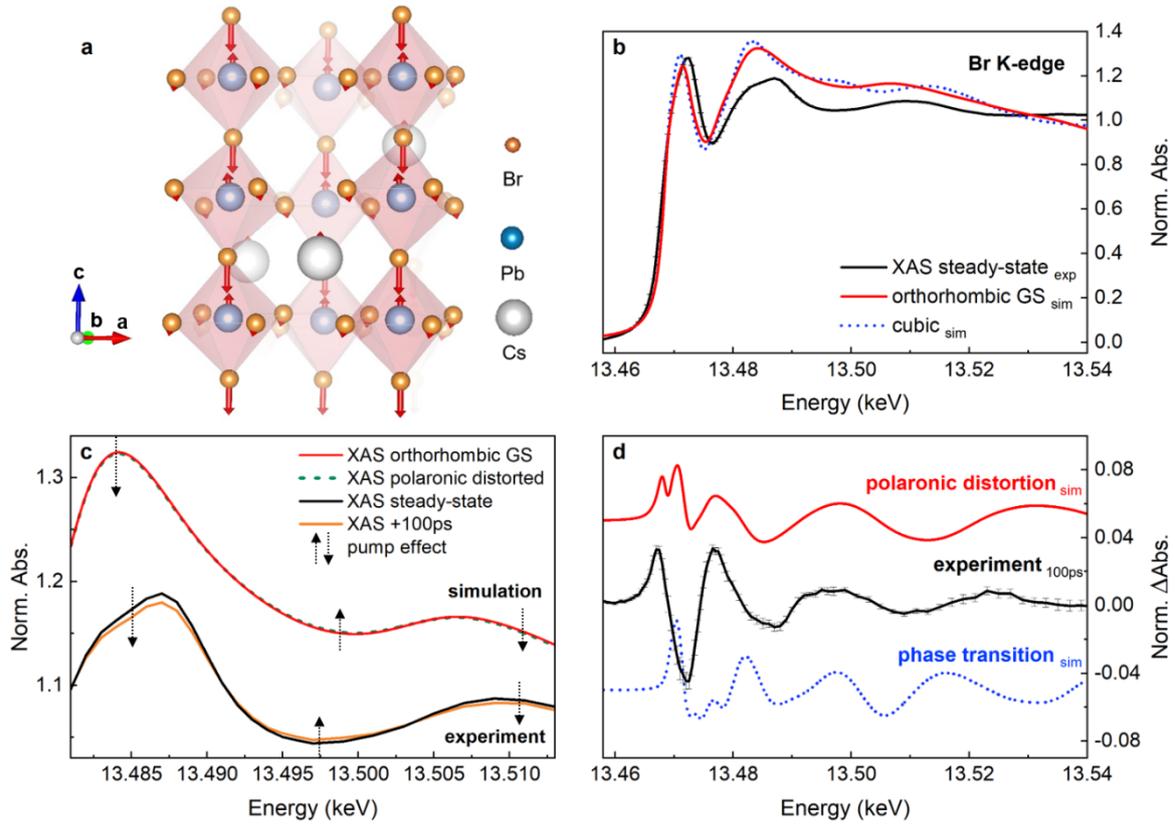

**Figure 4 | Theoretical ab-initio calculations of Br K-edge XAS spectra for the ground and structurally distorted states in CsPbBr$_3$.** (**a**) Schematics of the atomic displacements related with the 18 meV LO phonon mode. (**b**) Steady-state experiment (black, normalized for the last energy point), ground state orthorhombic simulation (red solid line, scaling factor x2550) and cubic simulation (blue dashed line, scaling factor x2550). (**c**) ground state orthorhombic simulation (red, scaling factor x2550), orthorhombic distorted along the 18 meV LO phonon mode (green dashed line, scaling factor x2550), unpumped (black, normalized for the last energy point) and pumped (yellow, 100 ps time delay, normalized for the last energy point) experimental spectra. (**d**) Experimental transient at 100 ps (black) and simulated pump-probe obtained as distorted orthorhombic minus pristine orthorhombic spectra (polaronic distortion, red) and cubic minus orthorhombic spectra (phase transition, blue dashed line). All spectra were scaled by the absolute area underlying the curves and the simulated pump-probe additionally multiplied by a factor 70 to enable the comparison with the experiment.

# Quantifying Photoinduced Polaronic Distortions in Inorganic Lead Halide Perovskites Nanocrystals Supplementary Information

**Table of Contents**

## 1. Samples and characterization

Table S1 summarizes specifications about the sample 1 employed in the Advanced Photon Source (APS) TR-XAS experiment and the sample 2 employed in the SuperXAS T-dependent XAS and XRD measurements. Figure S1 shows the absorption and emission spectra of the two samples, their transmission electron microscopy (TEM) images, and the high-angle annular dark-field scanning transmission electron microscope (HAADF-STEM) image of a single nanocrystal (NC).

| Name | Solvent | Concentration (mg/mL) | | Absorption (355 nm, 200 μm path) | Absorption Edge | Luminescence Maximum (FWHM) | Quantum Yield (%) | Particle Size (number of particles in TEM) |
|---|---|---|---|---|---|---|---|---|
| Sample 1 | Nanocrystals in toluene | 5.8 (in 75 ml toluene) | 23 (dried) | 1.7 OD | 508 nm | 516 nm (18 nm) | >90% (solution) | 10 ± 2 nm (61) |
| Sample 2 | Dry nanocrystals | - | - | - | 503 nm | 511 (22 nm) | 77.4 | 8.1±1.6 |

**Table S1.** Samples characterization for APS TR-XAS and SLS T-dependent XAS experiments.

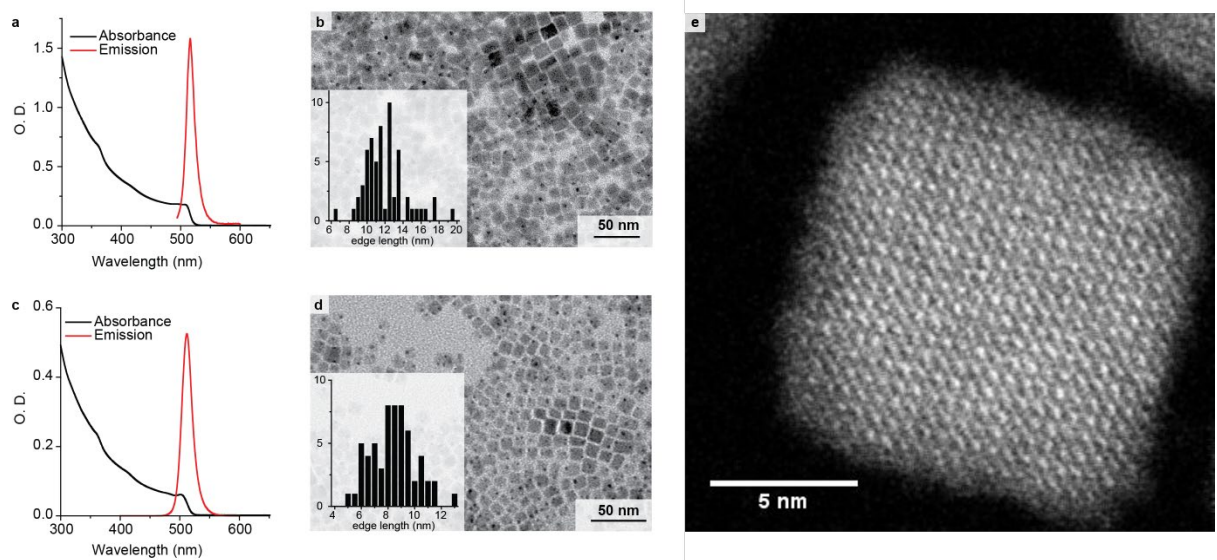

**Figure S1.** Absorption and Emission Visible Spectra and the TEM image of (**a**, **b**) sample 1, and (**c**, **d**) sample 2, respectively. The insets in (**c, d**) show the distribution of the average edge length (nm) of the cuboidal NCs, with the y-axis corresponding to the number of counts. (**e**) HAADF-STEM image of a single $CsPbBr_3$ NC.

## 2. Experimental methods

**TR-XAS:** The sample consisted of long-chain zwitterion-capped CsPbBr$_3$ perovskite cuboidal NCs with side length 10 ± 2 nm. Their preparation is described in [43]. The 5.8 mg/ml solution of CsPbBr$_3$ NCs dispersed in toluene was kept under stirring conditions and flowed through a sapphire nozzle to produce a 200 μm flat sheet jet, ca. 5 mm wide, ensuring that a fresh spot was excited at each pump pulse. The flat jet was oriented at 45° relative to the X-ray propagation direction. TR-XAS signals were collected at the 7ID-D beamline at the APS of the Argonne National Laboratory (U.S.A.), at the Br K-edge (13.450-13.570 keV) and Pb L$_3$-edge (13.000-13.140 keV), using a crystal diamond (111) monochromator and employing the ~80 ps (FWHM) pulses provided by the facility in 24-bunch mode, with a repetition rate of 6.52 MHz. The X-ray energy was calibrated using Pb foil. A high repetition rate Duetto laser system (10 ps FWHM pulses, pump wavelength 355 nm) was used to excite the sample at 8.8 mJ/cm$^2$. The laser repetition rate of 1.304 MHz was set to 1/5 of the 6.52 MHz repetition rate of the synchrotron source.

A nearly-collinear geometry between the X-rays (5 μm spot-size) and laser (45 μm x 50 μm spot-size) was employed, detecting the X-ray fluorescence signal in orthogonal geometry with Avalanche Photodiodes. Br K-edge and Pb L$_3$-edge spectra were collected both in total fluorescence yield (TFY), with two acquisition methods: analog averaging using a MHz boxcar average (Zurich Instruments) and single photon counting (SPC) using homemade FPGA-based gating electronics. Energy traces were recorded at 100 ps and 10.1 ns time delays at both edges. At every time delay, X-ray absorption spectra were recorded scanning each edge as a function of energy. A non-uniform grid of energy steps was chosen to optimize the acquisition times still preserving an energy resolution better than 0.5 eV, with finer steps in the edge region and wider steps in the above-edge region. Additional scans were retrieved at 163.5 ns time delay exploiting the higher repetition rate of the synchrotron (6.52 MHz) with respect to the pump (1.304 MHz), which implies that 5 X-ray beams consecutively probe the sample before a second pump pulse arrives.

Considering the small X-ray probe spot size and the average speed of the flat jet (~4 m/s), the second X-ray beam after the pump still probes a uniformly pumped region, arriving with an additional time delay of 153.4 ns (1/6.52 MHz). Thus, energy transients at 163.5 ns time delay were obtained subtracting the XAS signal of the unpumped sample from the XAS signal recorded with the second X-ray pulse after the pump, when the time delay between the pump and the first X-ray pulse was set to 10.1 ns.

A total accumulation time between 48 s and 66 s per energy point was employed for the averaged energy Br transients, such that the signal-to-noise ratio (SNR) of the measurements is mainly dictated by the signal strengths at the different time delays. At 100 ps the Br K-edge pump-probe signal at 13.472 keV is around 2% of the static signal. For the Pb L$_3$-edge, the pump-probe signal at 13.043 keV is around 0.4%. Thus, at the Pb L$_3$-edge, data acquisitions 5 and 9 times longer than Br transients were required to get a good statistics, respectively for 100 ps and 10.1ns/163.5ns. Despite the longer acquisition times, the Pb SNR remains lower than at the Br edge. The weak pump-probe response of the Pb centres can be partially attributed to the 1:3 stoichiometric ratio between Pb and Br sites in the sample and to the core-hole broadening of the heavy Pb element, which reduces the energy resolution of the measurement. Additionally, it could reflect the weaker response of the Pb sites to the optical pump. The static spectra in Figure 2(a, b) in the main manuscript were normalized for the last point of the post-edge, whereas Br transient spectra were scaled by the inverse of the absolute area underlying the curves, i.e. x155 (100 ps), x363 (10.1 ns), x976 (163.5 ns). The error bars were changed accordingly. Due to the higher SNR of Br transients and the correlation between the Br and Pb time responses (see main manuscript), the same scaling factors were used for the Pb spectra.

Time traces were obtained recording the spectral intensity at each X-ray absorption edge while scanning the delay between the optical pump and the X-ray probe. The traces were acquired over a range of 130 ns on Br K-main-edge (13.472 keV) and Pb $L_3$-main-edge feature (13.043 keV). The acquisition times were optimized to obtain comparable SNR levels between the two traces, requiring 36 s per time point at the Br edge and 6 times longer accumulation times at the Pb edge. The dataset in the main manuscript was analyzed following a global fit procedure, where the minimization of the residuals between the fitting curve and the Br and Pb datasets was performed simultaneously.

**Temperature-dependent XAS:** The temperature-dependent XAS and XRD measurements were performed at the SuperXAS beamline at the Swiss Light Source (SLS) of the Paul Scherrer Institute (Switzerland). The sample consisted in a powder of $CsPbBr_3$ NCs, allocated into a cell holder between two graphite layers with 0.254 mm thickness. The internal temperature of the cell was calibrated and monitored with a thermocouple. The XAS signal was acquired using a 5-element silicon drift detector (SDD) for fluorescence detection at a 90° geometry. The spectra were collected using a Si(111) monochromator and exploiting the complete filling pattern of the synchrotron delivered at 1.064 MHz repetition rate. Br K-edge (13.450-13.569 keV) spectra were recorded at the temperatures 25, 35, 40, 45, 55, 60, 65, 120, 140 °C, whereas smaller temperature steps were chosen for the Pb $L_3$-edge (13.000-13.140 keV) spectra, namely 25, 35, 40, 45, 55, 60, 65, 80, 100, 120, 140 °C.

XRD measurements were performed with the sample at 0° with respect to the incidence monochromatic X-ray beam (12.9 keV). The signal was collected using a Pilatus 100k detector (94965 pixels, 172x172 µm$^2$ pixel area) in transmission geometry. The sample-detector distance was 24.10 cm, with the detector laterally displaced of 4.53 cm, avoiding its exposure to the direct X-ray beam. XRD patterns were recorded at the same temperatures as XAS. The background XRD signal generated by the graphite sheets was isolated in a separate measurement.

### 3. Fluence scans data analysis

Figure S2 reports the fluence dependence of the pump-probe signal intensity at 100 ps time delay, in the range 0.5 to 12.4 mJ/cm$^2$. In the figure, the Br (13.472 keV, left axis in orange) and Pb (13.043 keV, right axis in blue) main-edges are displayed as a function of the 355 nm pump intensity. The linear fit of the Br data, with better SNR, is showed by the grey solid line, and overlaid to the Pb response.

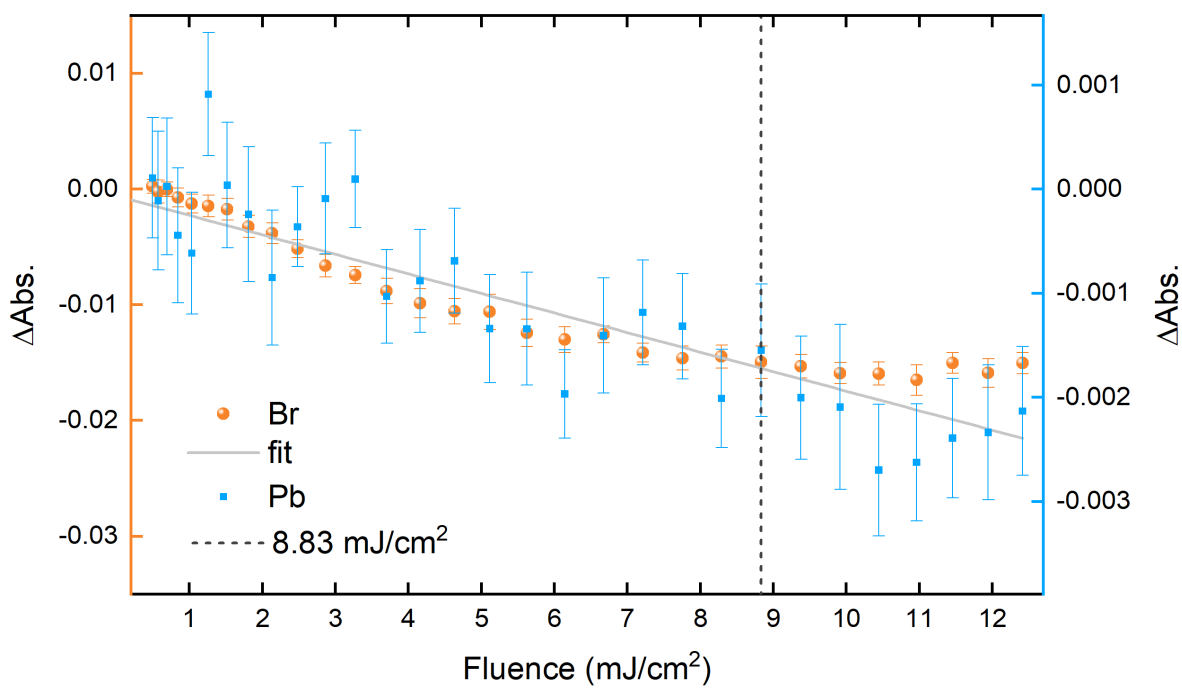

**Figure S2.** Pump-probe fluence dependence intensities at the Br (13.472 keV, left axis in orange) and Pb (13.043 keV, right axis in blue) main-edges. The error bars represent the standard error of the measurements. The dashed vertical line corresponds to the fluence employed in the pump-probe experiments. The grey solid line corresponds to the linear fit of the Br TR-XAS fluence-dependent signal in the linear regime.

Both signals show a linear regime up to ~10 mJ/cm$^2$, followed by the onset of a saturation plateau. Our pump-probe measurements were performed at 8.83 mJ/cm$^2$, approaching the highest fluence within the linear regime. Pb and Br data were recorded in TFY mode at 100 ps time delays. The TFY signals were divided by the incident X-ray flux, monitored with an ungated ion chamber. Laser-on scans were corrected by the offset with respect to the corresponding laser-off scans. The offset was determined as the average of the difference between laser-on and laser-off for the first two points in fluences (correction within the error bars amplitude). In order to compensate for possible concentration fluctuations among different scans, laser-off and laser-on scans were scaled by the mean intensity value of the corresponding laser-off scan. These two corrections were necessary to compensate the large variation in the static signal arriving at the detector in different scans. Laser-on and laser-off scans were averaged and their standard errors were computed as standard deviation scaled by the square root of the number of scans. Pump-probe scans were calculated as the difference of laser-on minus laser-off averages and the pump-probe error was computed through error propagation of the laser-on and laser-off errors.

**4. Energy scans data analysis**

XAS spectra were collected in TFY with the two acquisitions modes mentioned above, which required different data processing procedures. Analog signals averaged in the MHz boxcar were normalized by the incident X-ray flux, monitored with an ungated ion chamber, and the laser-off pre-edge offset was substracted from consecutive laser-off and laser-on scans. Each pair of offset-corrected scans was scaled by the laser-off edge integral and the pump-probe spectra were computed subtracting laser-off from laser-on scans. The pump-probe signal was obtained averaging the pump-probe scans and the corresponding standard error was calculated from the error propagation of the laser-on and laser-off spectra. The offset correction and edge-integral scaling were employed to correct the scan-to-scan fluctuations, probably due to jet instabilities.

The SPC laser-off and laser-on signals were corrected for the probability of having multiple photons hitting the detector and summed over all scans. The corresponding errors were computed as the square root of the total number of counts. Both the signal counts and their errors were scaled by the total incoming flux signal. Separate flat pre-edge offsets were removed from the laser-on and laser-off spectra and the signals were scaled by their edge-integral instensities. Their errors were also scaled by the same scaling factor (but no offset was subtracted, being the error proportional to the square root of the total number of counts). The pump-probe spectrum resulted from the difference of the laser-on and laser-off signals and its error was computed from the error propagation of the laser-on and laser-off errors. Br transients at 100 ps and 10.1 ns were obtained in TFY, averaging 9 and 11 scans respectively, whereas the spectrum at 163.5 ns was recorded in SPC, considering the counts accumulated in 8 scans. Pb energy transients were collected in SPC mode, summing the counts of 56 scans (100 ps) and 87 scans (10.1 ns and 163.5 ns).

### 5. Time scans data analysis

Both Br and Pb time traces were collected in SPC mode. The SPC counts were corrected for the probability of having multiple photons hitting the detector and a sum over all the scans was performed for laser-on and laser-off scans separately. The errors were computed as the square root of the number of counts. Both signals and errors were scaled by the total incoming flux intensity and separate offsets were subtracted from laser-off and laser-on spectra, retrieving an average zero intensity of the pump-probe signal at negative time delays. The pump-probe time trace was obtained from the difference of the laser-on and laser-off scans and its error was computed from the error propagation of the laser-off and laser-on errors.

### 6. Estimation of the number of photocarriers

Under the fluence conditions of the TR-XAS measurements, we estimate an initial number of photocarriers per NC of about 1000, as predicted by the formula:

$$N = F \cdot \mu \cdot V$$

where $F$ is the average 355 nm pump fluence absorbed by the solution in absence of light scattering (7.6·10$^{19}$ ph/m$^2$), $\mu$ is the intrinsic absorption coefficient of CsPbBr$_3$ NCs at 355 nm ($1.3 \cdot 10^7$ m$^{-1}$) [68] and $V$ is the average NC volume ($10^{-24}$ m$^3$). This average number corresponds to an upper limit and could considerably overestimate the actual number of charge carriers present in the system at the probed time delays. More precise calculations, that are beyond the scope of this study, should consider additional factors, such as light scattering, NC size inhomogeneity, and the statistical (Poisson) distributions of the absorbed photons in the NCs.

## 7. Computational methods and DOS calculation

In this section we discuss the computational approach used to simulate the XANES spectra at the Br K-edge. All the calculations were performed using the Quantum ESPRESSO distribution [44,45], an open-source tool for electronic structure simulations based on density functional theory (DFT) and plane-wave and pseudopotentials technique. The Perdew-Burke-Ernzerhof (PBE) functional [46] was used to describe exchange-correlation effects. The electron-ions interaction was modelled using ultrasoft pseudopotentials from the PS-library [47]. Semicore *3d, 5d,* and *5s* plus *5p* electrons are treated as valence states for Br, Pb, and Cs, respectively. The Kohn-Sham wave functions and the charge density were expanded in plane-waves up to a kinetic energy cut-off of 50 and 300 Ry, respectively. A threshold of $10^{-10}$ Ry/atom on the total energy is set to define the convergence.

The equilibrium structures for the orthorhombic and the cubic phases were taken from experiments [15]. The distorted orthorhombic structure was determined starting from the equilibrium one and following the atomic displacement corresponding to the phonon with the largest coupling to the electronic degrees of freedom. The lattice vibration eigenmodes were calculated using the matrix of force constant taken from ref. [37]. The most strongly coupled mode was identified by looking at the effective Fröhlich electron-phonon coupling [37] and corresponds to the Pb-Br stretching mode at 18.2 meV (see Fig. 4 in ref. [37]). The strength of the displacement was chosen in such a way to give a change of the order of $K_BT$ in the DFT energy of the orthorhombic unit cell (20-atoms cell). We verified that the results are independent on this parameter by performing additional simulations for the distorted orthorhombic structure doubling the strength of the displacement. No significant change in the XANES line shape was observed, other than minor amplitude changes which have negligible effects on the spectra differences.

The XAS cross section $\Sigma(\omega)$ is given by:

$$\Sigma(\omega) = 4\pi^2 \alpha_0 \hbar \sum_{f,\boldsymbol{k},\sigma} \left| \langle \varphi_{f,\boldsymbol{k}}^\sigma | \mathcal{D} | \varphi_{i,\boldsymbol{k}}^\sigma \rangle \right|^2 \delta\left(\varepsilon_{f,\boldsymbol{k}}^\sigma - \varepsilon_{i,\boldsymbol{k}}^\sigma - \hbar\omega\right)$$

where $\alpha_0$ is the fine structure constant, $\hbar\omega$ the energy of the incoming photons, $\varphi_{i,\boldsymbol{k}}^\sigma$ and $\varepsilon_{i,\boldsymbol{k}}^\sigma$ the wavefunction and energy of the initial state, and $\varphi_{f,\boldsymbol{k}}^\sigma$ and $\varepsilon_{f,\boldsymbol{k}}^\sigma$ the wavefunction and energy of the final state. $\mathcal{D} = \boldsymbol{e} \cdot \boldsymbol{r}$ is the operator describing the light-matter interaction in the dipole approximation, with $\boldsymbol{e}$ and $\boldsymbol{r}$ being respectively, the polarization vector of the incoming light and the electron position. The absorption cross-sections were calculated in the dipole approximation using an efficient Lanczos recursive method that avoids the expensive calculation of the empty states, as implemented in the XSpectra package [48,49] of Quantum ESPRESSO. For Br K-edge spectra, the initial state $\varphi_{i,\boldsymbol{k}}^\sigma$ is the 1s wavefunction of the emitting

Br atom and the finals states wavefunctions $\varphi_{f,\boldsymbol{k}}^{\sigma}$ and $\varepsilon_{f,\boldsymbol{k}}^{\sigma}$ were calculated within the so-called excited-electron plus core-hole (XCH) approach [56]. This approach requires a DFT calculation where one excited Br atom with a full core-hole in the 1s state is included in a supercell.

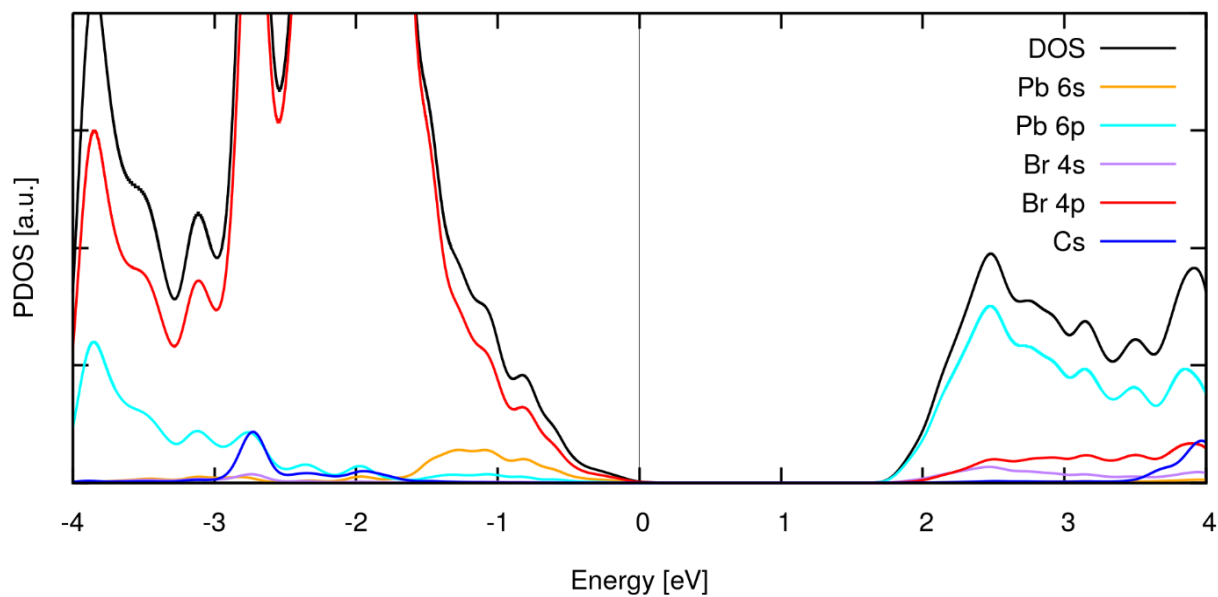

**Figure S3:** p-DOS in the band gap region. The zero of the energy is set at the top of the valence band, mainly constituted of 4p Br orbitals and with a small contribution of the Pb 6s orbitals. The conduction band is dominated by the Pb 6p orbitals, with minor contributions of the Br 4s and 4p orbitals. Cs orbitals do not significantly contribute to the DOS in the first few eVs around the band gap.

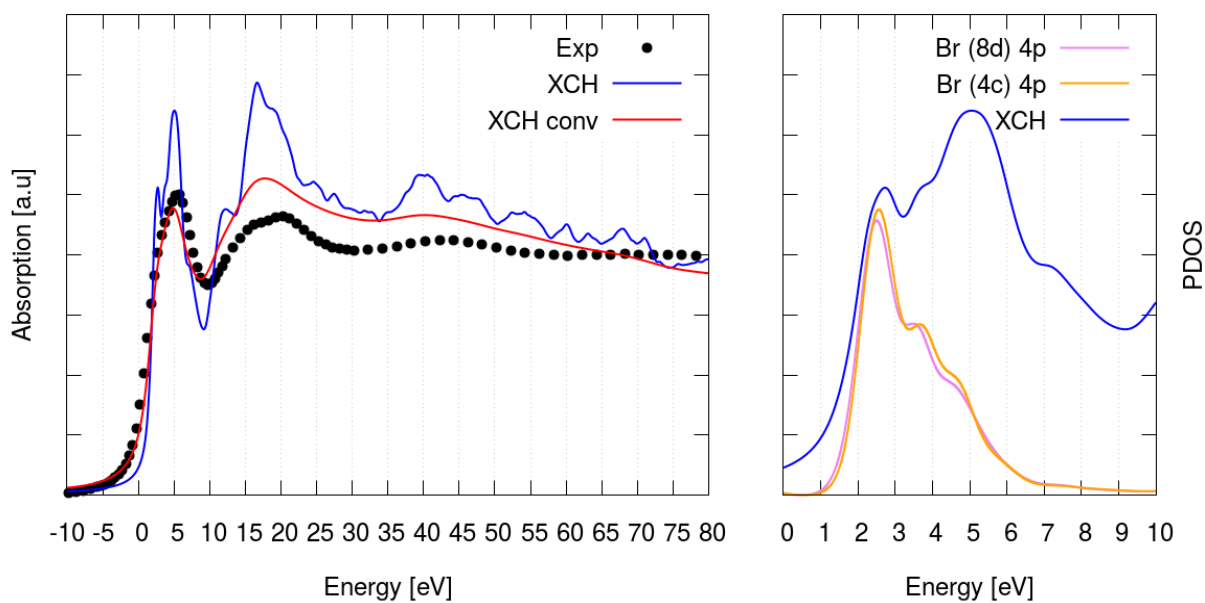

**Figure S4.** Left panel: the calculated Br K-edge XANES spectra (XCH) for the orthorhombic structure before (blue) and after (red) the Lorentzian convolution compared to the experimental data, which was rigidly shifted and aligned to theoretical spectra. Right panel: contribution to the p-DOS from the 4p states of the two nonequivalent Br atoms. The zero of the energy in the plots is set to the top of the computed valence band.

A pseudopotential for the Br atom in this excited configuration was generated with the LD1 module of the Quantum ESPRESSO package. An extra electron was added at the bottom of the conduction band leaving a charge-neutral excitation [56]. Separate XCH supercell calculations are performed for each non-equivalent Br atom in the structure and the final XANES spectrum is obtained from their average. A large enough supercell is needed to avoid spurious interaction of the excited atom with its replicas in periodic boundary conditions.

We verify that a unit cell containing 160 atoms, corresponding to a 2x2x2 supercell of the 20-atoms orthorhombic primitive cell, is enough to get converged results. To converge the Br K-edge spectra, the Brillouin zone of the supercell was sampled with a Γ-centred uniform grid of 3x3x2 **k**-points. The XAS intensities were averaged over three values of the polarization of the incoming light, namely along the [001], [010], [100] crystallographic directions. Pb $L_3$-edge calculations were not carried out due to the limitations of the XCH approach in describing holes with non-zero orbital momentum [50]. The averaged spectra were further convoluted with a Lorentzian having an energy-dependent broadening [69], which starts from 0.3 eV and reaches a maximum value of 6.00 eV, with an arctan-type behaviour and inflection point at 16 eV above the top of the valence-band. The projected density of states (p-DOS) for Br and Pb s, p orbitals are reported in Figure S3, showing a dominant contribution of the Br 4p orbitals and the Pb 6p orbitals in the valence and conduction bands, respectively. In Figure S4 the calculated Br K-edge spectrum is compared to the experimental data.

A good agreement with the experiment is achieved after the raw data (blue solid line) were convoluted with a Lorentzian function, as described above (red solid line). In the low energy region, the raw data shows two distinct peaks at 2.7 eV and 5.0 eV. After the convolution those merge into a single peak at ~5 eV. The shoulder just before this main peak is visible both in the theoretical and experimental spectra and originates from the first peak at 2.7 eV. These two peaks are directly connected to the p-DOS and come from Br 4p states, as highlighted in the right panel of Figure S4.

## 8. T-dependent XRD and XAS

### Thermal XAS at Br K-edge and Pb $L_3$-edge

Figure S5 shows the steady-state spectrum at 25 °C and the XAS differences between the temperature-dependent and the room temperature full spectra, recorded at three key temperatures for both Br and Pb edges.

XAS spectra were affected by a minor, systematic energy shift due to the backlash of the monochromator during the extensive energy scan from the Br to Pb edges and to the 12.9 keV energy of the XRD measurements. The source of this backlash was reliably identified because consecutive scans at the same edge did not display significant energy shift, even when varying the temperature. For each scan, the backlash was compensated for an optimal shift, determined through a minimization process as a function of the energy shift. We designed the optimization parameter as the absolute difference between the modulus of the area underlying the normalized first derivative of the considered spectrum and a reference spectrum. The 25 °C spectrum was used as reference.

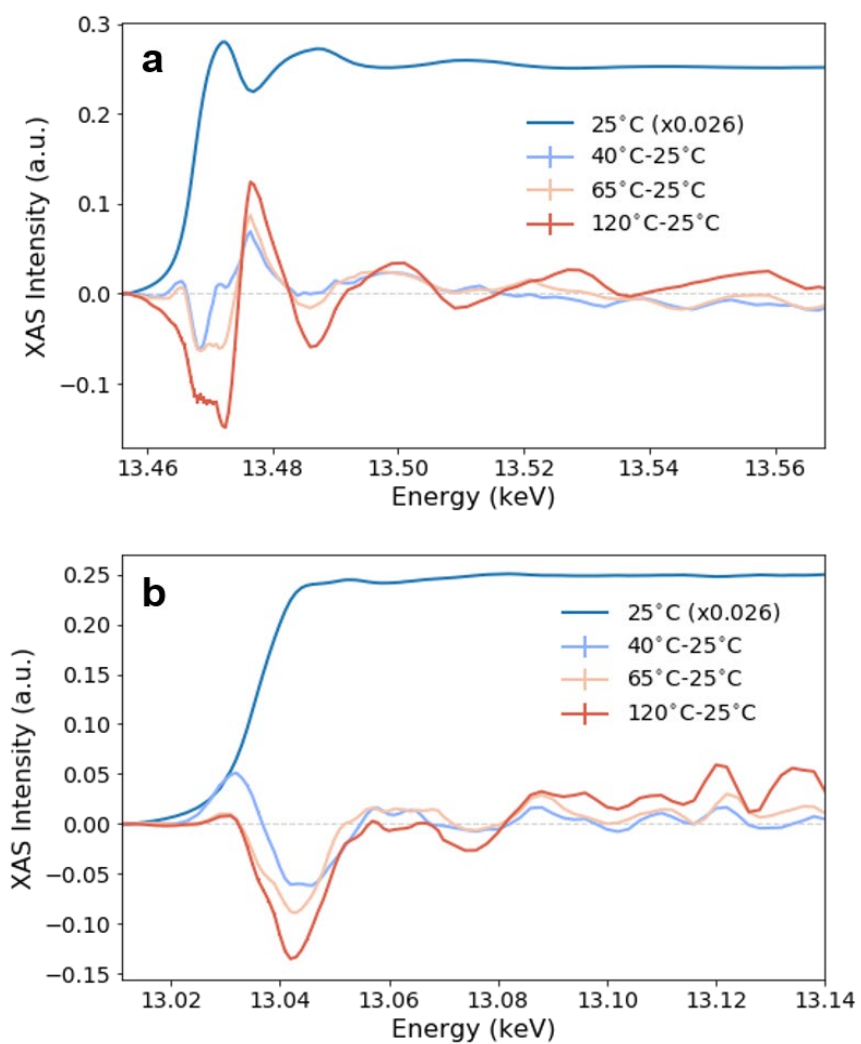

**Figure S5.** Differences between high-temperature XAS minus room-temperature XAS at (a) Br K-edge and (b) Pb $L_3$-edge, at three selected temperatures.

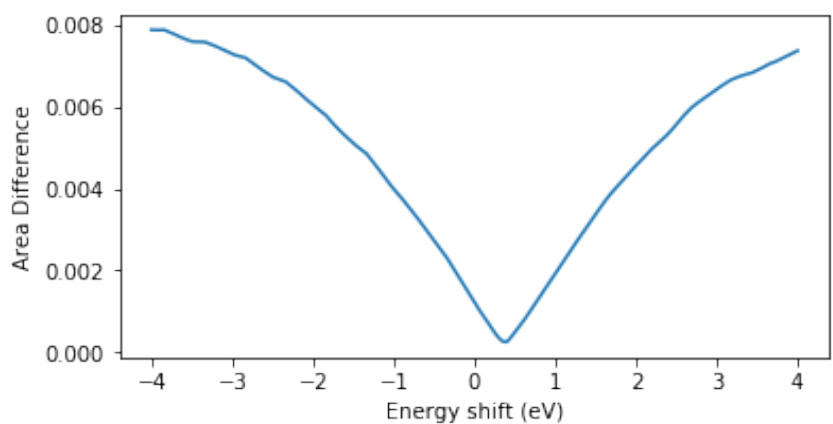

**Figure S6:** optimization parameter (absolute difference between the modulus of the area underlying the normalized first derivative of a pair of spectra) as a function of the energy shift for the Br K-edge scan at 45 °C. The 25 °C XAS spectrum was taken as reference spectrum.

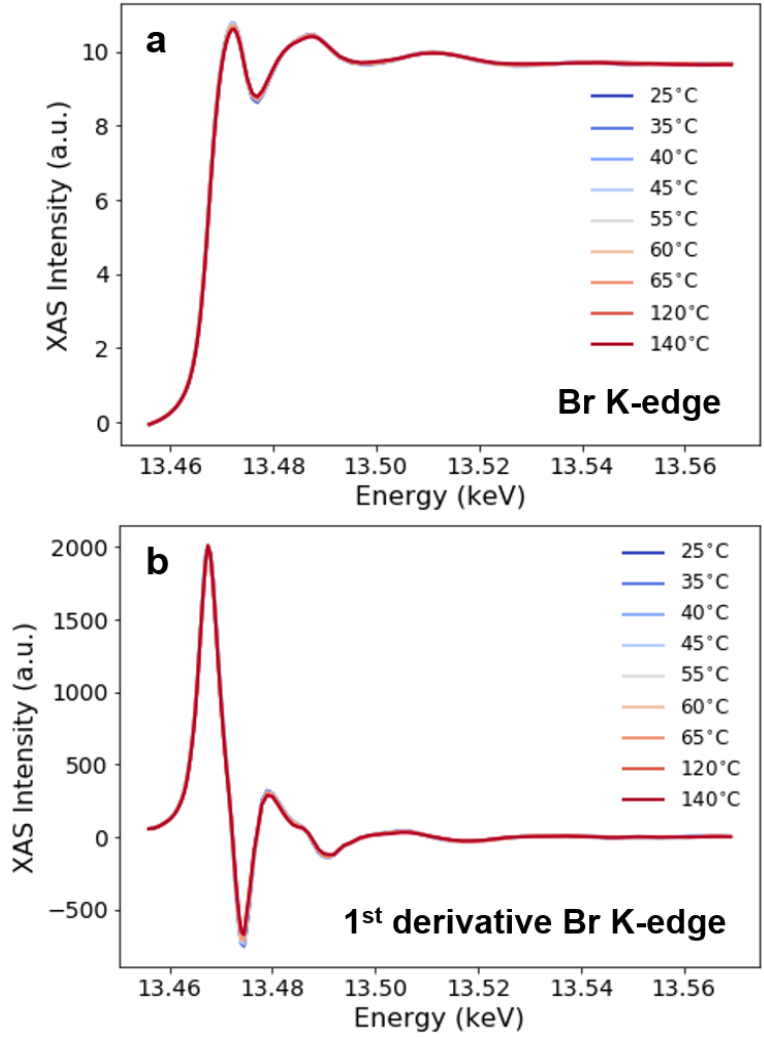

**Figure S7:** aligned and normalized Br K-edge (a) spectra and (b) first derivative with respect to the energy as a function of the temperature.

The same optimization procedure was employed to align the thermal dataset with respect to the steady-state Br K-edge and Pb L$_3$-edge XAS spectra collected in APS. For all scans, an optimal value for the shift was uniquely identified. In Figure S6 we report the typical profile of the optimization value as a function of the energy shift. The backlash correction could remove temperature effects on the XAS edge absolute energy position, as in the case of band gap temperature-dependence.

Indeed, assuming that the core orbital energies are not affected by changes in the sample temperature, the absolute position of the XAS spectra may change with the band gap, as it influences the energy difference between the valence and conduction bands. A correction was introduced following Varshni relation [70], estimating the temperature dependence of the system's band gap as:

$$E_g(T) = E_g(0) - \frac{\alpha T^2}{T + \beta}$$

In the case of CsPbBr$_3$ NCs, the values of α=-7.3·10$^{-5}$ eV·K$^{-1}$ and β=314 K were determined in a temperature-dependent photoluminescence study in the temperature range 80-550 K [71]. Under these assumptions, a blue-shift of the XAS spectra up to 6.5 meV can be expected heating the sample from 25 °C to 140 °C, much smaller than the backlash corrections applied to the temperature-dependent Br K-edge spectra (between 0.235 eV and 0.589 eV). For each temperature, the final spectrum was computed as the average of the backlash corrected spectrum and the same spectrum where Varshni correction was also included. The corresponding error bars were also calculated.

Consistently with the data analysis procedure of pump-probe spectra, a flat pre-edge offset was subtracted from each steady-state scan and the obtained spectrum was normalized by its edge integral. The resulting aligned Br K-edge spectra and their first derivative with respect to the energy are shown in Figure S7 as a function of the temperature. Temperature-dependent XAS differences were obtained subtracting the 25 °C XAS spectrum from each XAS spectrum in the temperature range 35-140 °C.

**Quantification of the photothermal effect**

We estimated the upper limit of the optically-induced temperature changes considering the energy deposited by the pump beam in the CsPbBr$_3$ lattice, in absence of cooling channels for the NC lattice on sub-100 ps time scales. Accounting for the exceptionally high photoluminescence quantum yield of our sample (~90%), the lattice temperature increase is mainly related to the deposition of the excess energy between the pump wavelength (355 nm) and the emission wavelength (516 nm). We estimate a value of ΔT~120 °C, following the equation:

$$\Delta T = \frac{(F_{abs} \cdot \alpha)}{(c \cdot d_{jet} \ast c_p)}$$

where we considered an effectively absorbed fluence $F_{abs}$ of 7.15 mJ/cm$^2$ (given the solution optical density of 1.7 O.D.), an energy-to-heat conversion α of 0.41 (which accounts for both the pump excess energy and the 10% of non-radiative recombination), a CsPbBr$_3$ concentration c of 5.8 mg/ml, a jet thickness $d_{jet}$ of 200 μm and the heat capacity $c_p$ value of 215 mJ/g·K, approximated to its lower limit with respect to the temperature [72,73]. Scattering of light from the liquid jet was neglected. Similar calculations were reported in ref. [33] to estimate the temperature increase in an ultrafast electron diffraction experiment on MAPbI$_3$ thin film.

Despite the instantaneous ΔT obtained in the calculation is non-negligible, we can exclude the presence of dominant photothermal effects in our measurements, based on the expected time scales for the cooling process and the spectral changes associated to it (see main manuscript).

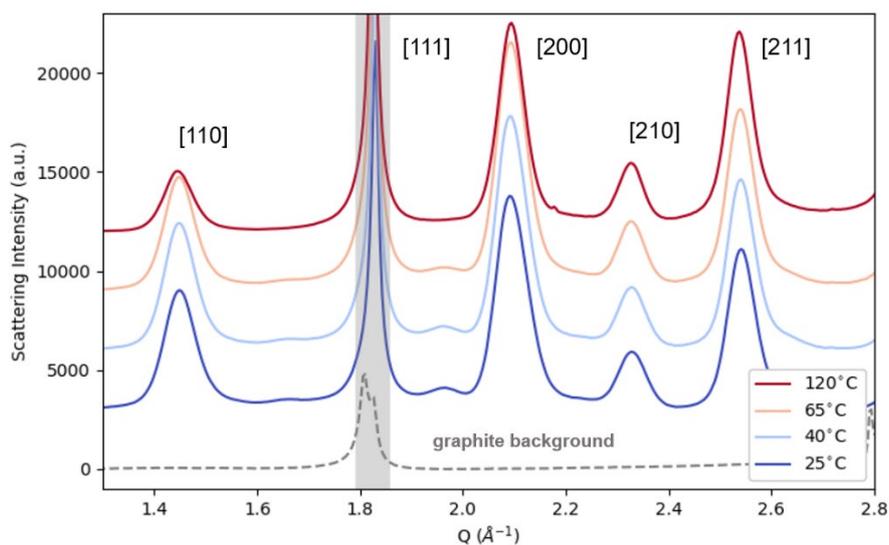

**Figure S8.** XRD in the 1.3-2.8 Å$^{-1}$ Q range at four selected temperatures. For the 120 °C pattern, the peaks are labeled accordingly to the high-temperature $Pm\bar{3}m$ cubic phase reported in the literature [36]. An arbitrary offset was introduced for each spectrum to better resolve the spectral evolution with the temperature. The dashed grey curve corresponds to the background originated by the graphite sheets of the thermostated cell, which signal is superimposed to the [111] CsPbBr$_3$ diffraction peak.

**T-dependent XRD data analysis**

T-dependent XRD data were collected using a Pilatus 100k detector, with an incoming X-ray energy of 12.9 keV, in order to monitor the sample integrity in the thermal study. The 2D images were integrated using the python function pyFAI.AzimuthalIntegrator. The sample position with respect to the detector was determined calibrating the measured graphite 2D pattern with the graphite literature data reported in the "pyfai calibration" open source program. The rotation parameters were optimized in the program in order to correct the circles of the 2D maps into straight lines.

The 1D XRD patterns were thus obtained performing a vertical binning of the rectified 2D images. A common linear backgorund was subtracted by all curves. It was determined by minimizing the residuals between the 35 °C XRD background pattern and a linear function used in the fitting. The pixel-to-scattering vector conversion was performed using the following relations:

$$l = p \cdot \Delta$$

$$\theta = \tan^{-1}\frac{l+\alpha}{L}$$

$$d = \frac{\lambda}{\sin\theta}, S = \frac{2\pi}{d}$$

Where p is the pixel number corresponding to the peak position, Δ is the pixel size, L is the sample-detector distance, α is the distance between the transmitted X-ray beam and the detector in the plane orthogonal to the X-ray beam, λ is the 12.9 keV X-ray energy in wavelength. A least-square fitting was used to refine the measured L and α quantities in order to match the peak position of the XRD pattern generated with VESTA for a CsPbBr$_3$ orthorhombic unit cell. A Gaussian broadening was applied to VESTA predictions to better resemble the experimental conditions. In Figure S8, XRD profiles at selected temperatures are shown over the Q range between 1.3 and 2.8 Å$^{-1}$. The prominent peak at 1.83 Å$^{-1}$ comes from the background signal of the thin layers of graphite encasing the CsPbBr$_3$ sample.